# Practitioners' Perspectives on Change Impact Analysis for Safety-Critical Software – A Preliminary Analysis

Markus Borg[1], José-Luis de la Vara[2], and Krzysztof Wnuk[3]

[1] SICS Swedish ICT AB, Lund, Sweden
`markus.borg@sics.se`
[2] Carlos III University of Madrid, Spain
`jvara@inf.uc3m.es`
[3] Blekinge Institute of Technology, Karlskrona, Sweden
`krzysztof.wnuk@bth.se`

**Abstract.** Safety standards prescribe change impact analysis (CIA) during evolution of safety-critical software systems. Although CIA is a fundamental activity, there is a lack of empirical studies about how it is performed in practice. We present a case study on CIA in the context of an evolving automation system, based on 14 interviews in Sweden and India. Our analysis suggests that engineers on average spend 50-100 hours on CIA per year, but the effort varies considerably with the phases of projects. Also, the respondents presented different connotations to CIA and perceived the importance of CIA differently. We report the most pressing CIA challenges, and several ideas on how to support future CIA. However, we show that measuring the effect of such improvement solutions is non-trivial, as CIA is intertwined with other development activities. While this paper only reports preliminary results, our work contributes empirical insights into practical CIA.

## 1 Introduction

Safety-critical software systems evolve during their lifecycle. As changes are made to the systems, change impact analysis (CIA) is needed, defined as "identifying the potential consequences of a change in a system, or estimating what needs to be modified to accomplish a change" [2]. CIA is essential for safety assurance, and it is indeed prescribed by safety standards, e.g. IEC 61508 states that "if at any phase of the software safety lifecycle, a modification is required pertaining to an earlier lifecycle phase, then an impact analysis shall determine (1) which software modules are impacted, and (2) which earlier safety lifecycle activities shall be repeated."

CIA is often a difficult task in practice due to the size and complexity of safety-critical systems [2, 7]. Inadequate CIA has further been among the causes of accidents and near-accidents in the past [10]. Industry can clearly benefit from new CIA tech-nology and knowledge to more cost-effectively perform this safety assurance activity, enabling better risk avoidance and mitigation. Such technology and knowledge must be linked to current practices and targeted at meeting industry needs and expectations.

Despite the importance of CIA for safety-critical systems, the current knowledge about the state of the practice is limited. We are not aware of any publication that has studied the CIA activity in depth. The available knowledge is based on studies that (1) have dealt with non-safety-critical systems, (2) have analyzed data from past projects, (3) have not focused on CIA, or (4) have surveyed practices for safety-critical systems from a general perspective. For example, Rovegård *et al.* [14] interviewed software practitioners to analyze CIA issue importance, whereas Borg *et al.* [3] studied past issue reports of an industrial control system. Some insights have been provided in studies on e.g. the alignment of requirements with verification and validation [1] and on traceability [13]. Regarding the surveys, Nair *et al.* [12] studied safety evidence management practices, including certain aspects related to change management, and de la Vara *et al.* [7] conducted a survey on safety evidence CIA to explore the circumstances under which it is performed, the tool support used, and the challenges faced.

We have conducted an industrial case study on CIA for safety-critical systems in practice, particularly exploring engineers' views on the work involved. The context is a distributed development organization offering industrial control systems to a global market. We interviewed 14 engineers in two units of analyses, constituted of two teams located in Sweden and India, respectively. This paper reports a preliminary analysis covering a subset of the interview guide.

Our long term goal is to support architectural decision making when evolving cyber-physical systems, an endeavor in which the CIA is fundamental. As a step in this direction, we explore three research questions: RQ1) How extensive is the CIA work task?, RQ2) What are the engineers' attitudes toward CIA?, and RQ3) How could CIA be supported? By better understanding CIA in a particular case, we can take steps toward understanding how previous knowledge could be stored to support decision making in software evolution, in line with our previous work on traceability reuse [3] and knowledge repositories [6].

The rest of the paper is structured as follows: Section 2 presents the case, and Section 3 describes the research methodology. We report our results and discuss their implications in Section 4, and Section 5 concludes the paper and outlines future work.

## 2    Case Description

The case company develops safety-critical industrial control systems. The system under study is has evolved since the 1980s and needs to fulfil the IEC 61511 standard via Safety Integrity Level 2 certification, according to the IEC 61508 standard. The developed software must be of high quality and therefore all changes to source code need to be analyzed prior to implementation. Moreover, detailed system documentation is created and mapped to the vertical abstraction layers in the V-model. The projects follow a rigid development process with hundreds of collaborating engineers distributed globally. The code base is over one million lines, dominated by C/C++ and some newer extensions in C# or VB.

Prioritized features originating from various customers (and sometimes pre-ordered feature requests) are incrementally added and extensively tested. When developing new features and also when fixing issues to previously delivered features, several changes are made to the source code. When the development is over, the development organization needs to present a safety case for an external assessor, illustrating that the system is acceptably safe for a given application in a given operating environment. The set of CIA analyses is a crucial component of the safety case. Therefore, the safety engineers at the case company have developed a semi-structured CIA report template, cf. Table 1, to support the safety case in relation to the IEC 61508 safety certification. The developers use this template to document their CIA before committing source code changes.

Currently there is limited tool support available for CIA in the organization. The CIA process is tightly connected with the issue management process, as all changes to formal development artifacts require an issue report in the issue repository. All completed CIA reports are stored in the issue repository as attachments to issue reports. Developers typically access the issue repository using a simple web interface.

Table 1. CIA template used in the case company, adapted from Klevin [8].

| |
|---|
| Q1) Is the reported problem safety critical? |
| Q2) In which versions/revisions does this problem exist? |
| Q3) How are general system functions and properties affected by the change? |
| Q4) List modified code files/modules and their SIL classifications. |
| Q5) Which library items are affected by the change? (e.g., library types, firmware) |
| Q6) Which documents need to be modified? (e.g., reqts. specs, architecture) |
| Q7) Which test cases need to be executed? (e.g., design/functional/sequence tests) |
| Q8) Which user documents, including online help, need to be modified? |
| Q9) How long will it take to correct the problem, and verify the correction? |
| Q10) What is the root cause of this problem? |
| Q11) How could this problem have been avoided? |
| Q12) Which requirements and functions need to be retested by test organization? |

## 3   Research Methodology

We conducted a multiple unit industrial case study since the studied phenomenon could not be separated from its context [15]. The case under study is the CIA activity in the development organization described in Section 2. Two development teams constitute the units of analysis, referred to as Unit Sweden and Unit India, respectively. Figure 1 shows an overview of the study.

Four researchers iteratively (1) *designed* the case study and documented it in a case study protocol. All the steps in the design were reviewed by senior researchers. We constructed an interview guide (available online[1]) for semi-structured interviews to be able to ask both close and open-ended questions. We asked open questions in the beginning and end of the interviews, in line with the time glass interview model

---
[1] http://serg.cs.lth.se/fileadmin/serg/ImpRec_EvalStudy/ImpRec_InterviewGuides.pdf

[15]. This paper reports only on an analysis of a subset of the questions asked. Table 2 maps the RQs to specific parts of the interview guide.

The data collection consisted of (2) *interviews* in Swedish or English. For confidentiality reasons, a single researcher conducted all interviews. The same researcher (3) *transcribed* all interviews word by word and sent them back to the interviewees for (4) *validation*. We interviewed 14 engineers of whom 10 are developers that write source code and its documentation. More specifically, we interviewed one R&D manager, one safety engineer, three senior developers (incl. the team leader), and one junior developer in Unit Sweden, and one product manager, one technical manager, four senior developers (incl. the team leader), and two junior developers in Unit India.

As a preliminary (5) *analysis* step, the transcripts were copied into a spreadsheet divided according to the interview guide. Longer answers were divided into smaller chunks. The spreadsheet was then (5.1) *cleaned* to remove obsolete and unimportant pieces of spoken language.

The iterative (5.2) *coding* process started by highlighting key statements, to establish a quick overview of the data and support subsequent data navigation. We then applied the following coding schemes: Pre2-a followed a two dimensional axial coding [16] with "connotation to CIA" vs. "importance of CIA" coded into the interval [-2, 2] to express positive/negative connotation and importance, respectively; Pre2-b used predefined closed codes: daily, weekly, bi-weekly, monthly, and rarely; Pre2-c was coded onto a timeline, expressing either point estimates or (min.-avg.-max.) intervals; Pre2-d employed open coding that evolved into 12 codes: test cases, documents, information overload, prolonged time, motivation, confidence, requirements, time estimation, avoidance, root cause, system version, and conformance; Pre3 used the interval [-2, 2] to signal whether the proposed metrics were indicative of time needed, and difficulty to conduct, a CIA; Pre4 used open coding that developed into: tool, training, template, search, traceability, and reviews. The coded interviews were then (5.3) *analyzed* to detect patterns and draw conclusions based on the qualitative data.

Finally, we (6) *report* our results in this publication. To preserve the confidentiality of the interviewees, we do not provide full traceability from the answers. We use the labels engineer/developer and junior/senior when needing to be more specific.

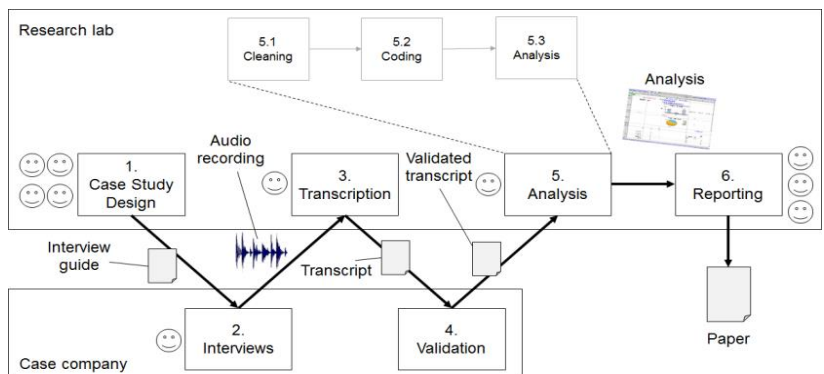

Figure 1. Overview of the research process. Smileys depict the number of researchers involved in the various steps.

Table 2. Breakdown from RQs to specific parts of the interview guide.

| RQ | ID | Part of the interview guide[1] |
|---|---|---|
| RQ1. How extensive effort do engineers spend on CIA? | Pre2-b | Do you conduct CIAs daily/weekly/monthly? |
| | Pre2-c | How much time do you spend on a CIA? |
| | Pre2-d | What are the greatest CIA challenges? |
| RQ2. What are the engineers' attitudes toward CIA? | Pre2-a | What is your general opinion about the CIA work task? |
| RQ3. How could CIA be supported? | Pre4 | What kind of support would you like to have access to when conducting CIAs? |
| | Pre3 | We have collected data from your CIA history. Could you please comment on the metrics?<br>- Pre3-b: Do the numbers reflect how long it takes to conduct a CIA? [TIME]<br>- Pre3-c: Some CIAs have been modified. Does this mean they were harder? [MODS] |

## 4 Results and Discussion

### 4.1 RQ1: Extensiveness of the Change Impact Analysis Task

To understand how extensive the engineers consider the CIA work task to be, we investigate: 1) how frequently engineers perform CIA, 2) how much time is needed to complete a CIA, and 3) what engineers consider as the major CIA challenges.

**Change Impact Analysis Frequency.** All the interviewees experienced with CIA stressed that the frequency of conducting CIA varies much, from daily to monthly CIAs, depending on the development phases. Four interviewees reported that they sometimes conduct CIA on a daily basis, and other four estimated that they do it weekly during the most intensive periods. Four other interviewees explained that their CIA intensity goes down to monthly during certain periods, while three interviewees answered bi-weekly CIAs at slow times. Five interviewees shared only estimates of their average CIA intensities: four reported weekly and one stated daily. A senior engineer in Unit Sweden estimated that a typical developer conducts 20 CIAs related to bug corrections per year. A senior engineer in Unit India stated a higher estimate: "in the thick of a project, a developer will do [a CIA] almost every day". We conclude that the engineers estimate the average intensity to be one CIA per week, considering variations due to cyclic development phases and parallel projects.

    The main reason for the variation is the stage-gate development process employed by the company; at the initial stages of a project, when the bulk of the new development is conducted, there is considerable source code churn, and changes are not managed on the level of individual issues. A senior engineer explained "we package new development as generic items in our issue tracking system, and then we conduct one comprehensive CIA". Once a project reaches the "code complete" milestone and the

formal verification phase is initiated, the goal is to stabilize the quality of the system, and all changes after this stage are considered bug corrections. The change management process then increases its granularity to individual bug resolutions.

**Change Impact Analysis Effort.** We asked the interviewees how much time they invested in a CIA, by providing the minimum, average, and maximum time needed. For the four non-developer interviewees, we instead asked them to approximate the time developers spend on CIAs. Four interviewees reported only an average CIA effort, i.e., a point estimate. Figure 2 presents an overview of the collected data. Similarly to the frequency results, the time required to conduct an individual CIA varies substantially. A senior developer stressed that the effort needed depends on the complexity of the involved component, and another senior developer expressed that it also depends on the structure of the corresponding documentation.

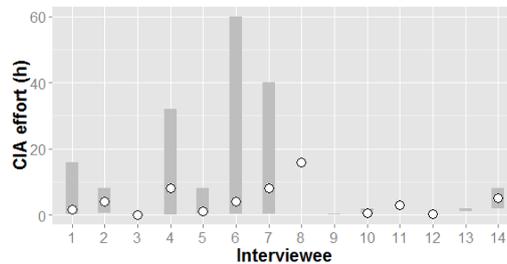

Figure 2. Interviewees' estimates of CIA effort. Bars depict minimum and maximum effort, circles show the average. Interviewees 5-14 are developers.

Three interviewees estimated that the average CIA requires 1-2 hours, four interviewees answered 4-6 hours, and three interviewees reported 1-2 days. On the contrary, two interviewees answered that they complete the average CIA in less than an hour, more specifically 30-40 min and 5 min, respectively. A senior engineer shared his rule of thumb regarding issue management and CIA: "a normal issue, or slightly more challenging than normal, takes in total roughly a man week to resolve. I estimate the CIA to require 10% of that time, 4 h. /…/ Then there are much faster issues that are resolved in a man day. 10% of that means about 1 h for the CIA".

Eight of the interviewees claimed that the minimum time required to complete a CIA is 30 minutes or less, of which two answered 5 minutes or less. Two senior developers in India reported that quick CIAs still require an hour or two, motivated by "there are 13 questions to correctly answer" and "if you fill it in honestly, without copy-paste". Regarding the maximum time, four interviewees claimed 1-2 days, two answered a week, and a junior developer instead expressed "several months of calendar time". Two senior developers in India shared contrasting views, claiming that the most time-consuming CIAs still could be finished in 15-20 min.

We noted three main explanations for the reported variation. First, it is *hard to detach the CIA from the general issue resolution, thus the interviewees interpreted the questions differently*. The exploratory work involved in reproducing and understanding an issue is often intertwined with the CIA. To understand an issue well enough, developers must sometimes set up a specific test environment, stepping through source code, etc. As a senior developer put it: "there is considerable exploratory work, and it would require almost as much time without the formal CIA". Second, the *complexity among the parts of the system is not constant*. The same interviewee considers issues related to the embedded environment to be the most complex, and maintenance issues on isolated components to be simple. Quick CIAs occur when an issue needs to be corrected in several system versions, and parts of the CIAs can be reused. Third, *experienced developers conduct CIAs faster than novices*. Three junior developers reported relatively high estimates, and two senior engineers stressed the importance of experience and system understanding for successful CIAs.

**Major CIA Challenges.** We received 25 challenges from 14 interviewees that cover both general and specific aspects, presented in Table 1. Note that Table 1 comprises also questions that deal with issue management in general, e.g. Q9-Q11, as reflected by some interviewees' answers.

The most frequently mentioned general challenge was related to *motivation*, i.e., understanding why comprehensive CIAs are part of the process and recognizing their value. The difficulty to appreciate the CIA activity was primarily expressed in the Swedish unit of analysis, by both seniors and juniors alike; seniors struggling more with motivating others, whereas juniors talked about motivating themselves. A senior engineer explained: "my main challenge is to explain and motivate why we do CIA, because if you know why you have to do it, you accept it. But we must continuously remind ourselves why we do it." and "if we get the developer to see that the CIA is very good, it helps me in my work /…/ That's what we´re aiming for".

The second most commonly mentioned general CIA challenge is related to *information overload*. The three most senior engineers in the study, explained that obtaining a system understanding is hard due to the complexity. Apart from the source code, there are numerous documents describing the system. The sheer number of software artifacts contests the system overview, and makes traceability information highly complex. A senior engineer stated: "finding the right information has historically been the major challenge /…/ In principle you had to hunt down key people and ask for documents and dependencies, and you didn't necessarily get the answer".

Three additional general CIA challenges were mentioned. First, sometimes developers need to *update previous CIAs that were conducted a long time ago* (half a year before according to a junior developer). Returning to old issues is difficult and requires considerable time and effort. Second, one interviewee said that his major challenge was to *trust his own CIAs*, i.e., establishing confidence in the answers to the questions in Table 1. As he explained it: "I'm not sure whether my answers are true or not, I cannot evaluate it. I am the judge." Finally, one senior interviewee expressed that the major challenge was that the *developers do not follow the CIA guidelines*.

Interviewees reported three CIA challenges specific to the questions in Table 1, all of them reported by two or three interviewees: 1) *selecting which test cases should be executed* to verify the changes (Q7), 2) *understanding which requirements are affected* by a change (related to Q6 and Q12), and 3) *reporting which documents need to be updated* to reflect the change (Q6). A senior developer explained that "the question about test cases is supposed to cover both directly and indirectly affected test cases, and the indirect ones are quite difficult." Regarding the requirements, another senior developer said: "Requirements traceability is difficult. We do not have requirements that cover all aspects, and it is hard for developers to stay on top of all requirements. [As developers] we don't work continuously with the requirements."

Concerning the challenges that deal with issue management in general, the interviewees in Unit India reported four major challenges. Three interviewees explained that *finding the root cause of an issue* (Q10) is a major challenge. Two interviewees pointed out suggesting *how the issue could have been avoided* (Q11) as particularly difficult, and two others highlighted determining *which system versions are affected by the issue* (Q2). Finally, a senior developer in Unit India considered *estimating the resolution time* (Q9) as the major challenge. Our interviews suggest that developers answer the three questions Q9-Q11 with different levels of ambition, and two interviewees from Unit India indicated that they invest considerable effort answering these questions. Concerning determining affected system versions, a senior developer clarified: "sometimes you study the source code, sometimes you have to run tests in our lab [on multiple versions] /.../ This is not hard, but time-consuming."

Based on our interviews, it appears that determining how a change impacts the product source code is less of a challenge than determining impact on non-code artifacts, e.g., requirements, specifications, and test cases. Considerable research effort has been directed at CIA on the source code level [11], but neither junior nor senior engineers discussed source code in relation to the major challenges involved in CIA. Our results instead support conclusions from Lehnert [9] and de la Vara *et al.* [7], i.e., there is a need for CIA research that considers different artifact types.

### 4.2   RQ2: Connotations of Change Impact Analysis

We explored the engineers' connotation to "change impact analysis", i.e., the emotional association carried by CIA, in addition to the explicit or literal meaning (the denotation). Furthermore, we map the connotation to the impression of how important CIA is to the individual interviewee, as presented in Figure 3.

The interviewees' attitudes toward CIA represent all quadrants in Figure 3. No interviewees expressed very strong positive or negative connotations to "change impact analysis". Most answers are balanced and we note that a majority of the engineers consider their CIAs important. Several interviewees shared positive associations, e.g., "a healthy sign" and "shows that we do complex software engineering". Also, several interviewees had a neutral connotation: "no values on a personal level" and "just part of the job". On the other hand, two interviewees expressed negative connotations: "a too heavy construct in our organization", "unfortunately very rigid activity", and "could be done in a better way".

Considering the interviewees' perceived importance of the CIA activity to their work, all levels are covered. Responses range from "for some issues it is just worthless stuff, done for the process" to "professionally, it is very fundamental" and "side effects are extremely important in our complex product". We identified no indications of attitude differences between the two units of analyses. Regarding the level of sen-iority, there is a slight tendency of senior engineers considering CIA to be more important than juniors. This is no surprise, as the seniors have seen more cases of source code changes causing side effects, especially from earlier development work when the CIA was less formal, as explained by one senior engineer "we have seen many cases when fixes introduced bugs". On the other hand, lack of experience can also induce a positive attitude, as expressed by one junior developer in Unit India: "This is my first company. Whatever I see, I feel is good".

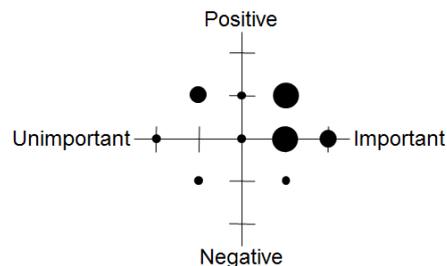

Figure 3. Interviewees' attitudes toward CIA. The y-axis shows the general connotation of CIA, whereas the x-axis depicts how important interviewees consider the CIA.

### 4.3  RQ3: Supporting Change Impact Analysis

**Envisioned Change Impact Analysis Support.** As one might expect when asking software engineers for solutions to support CIA, *the most frequent answer was tool support*. Three interviewees suggested increasing the level of automation in CIA by introducing some kind of tool. Interestingly, all these interviewees are seniors, and all discuss approaches to identify impact beyond the source code level. A possible explanation is that senior engineers are well-aware of the large efforts on establishing traceability in the organization, and that they consider the investment to be underutilized. One interviewee explained how a tool could iteratively follow traces through the system, from an input source code file or module, to a design description and its corresponding design tests, then continue up the abstraction layers to functional specifications and all the way up to the requirements. This could allow test identification directly from requirements, as a tool-oriented solution that would close the often challenging gap between requirements and test cases [1]. Another suggestion given by a recently employed senior developer was that "it is strange to attach a CIA report as a free-text string in a third party tool. In my previous company, this would have been done in an internally developed tool to better control the input [...] It would have been much easier to manage and interpret the data if the collection was more controlled." Indeed other companies use internal tools for CIA of safety-critical systems [7].

Two senior interviewees proposed *changes to the CIA template* in Table 1. A senior engineer considered security to be of critical importance to the future of the organization, and requested *adding explicit security questions* to the template. The other interviewee would like to see the question about *root cause analysis to be more elaborate* (Q10). The same interviewee also advocated *using more than one version of the CIA template, adapting the questions to the issue at hand*, to address the impression of having a too rigid CIA process. Another suggestion mentioned by two interviewees, a junior and a senior engineer, is to *increase the training of new employees on the specifics of CIA in the case company*, motivated by: "a new guy needs to be thoroughly trained on these questions /…/ I see newcomers just filling in things for the sake of moving the issue through the process".

Individual interviewees proposed additional improvement suggestions. First, *the search functionality in the issue tracking system should be improved*. Although no interviewee reported the current issue tracking system to be a major challenge to CIA, several mentioned that the tool was old and not user friendly. Another considerable limitation is a lack of full-test searchers in the issue tracker. Our previous work explored the potential of introducing state-of-the-art search technology to index full-text descriptions of issue reports [5]. Second, one senior engineer considered the source code and the documents to be two discrete information spaces with few connections. Despite all efforts to maintain traceability at the company, the interviewee wanted to *improve bi-directional connections between documents and source code*. Third, one junior developer thought that the CIAs "are not taken in a serious way", and proposed *introducing CIA reviews* before submitting them, as before committing code.

**Measuring Change Impact Analysis Support.** Prior to the interviews, we explored the CIA history stored in the issue tracking system, and constructed two measurements: TIME and MODS. We calculated these measurements for the relevant interviewees and presented the results during the interviews.

The first proposed measure, TIME, is the *time between a developer is assigned an issue and the first CIA report is submitted*. TIME thus targets CIA effort, but no interviewee considered the measure to be correlated with the time it takes to actually conduct the CIA. Six interviewees directly rebutted the measure, and two did not know how to interpret it. The most skeptical views explained: "it's definitely a question of priorities", "I work on several parallel products, and that measure can be anything", and "you have to measure when I start making related changes". We conclude that *TIME is a too confounded measure* to be used for evaluating solutions that aim at decreasing the time needed for CIAs.

The second proposed measure, MODS, is the *number of modifications on a CIA report after its first submission*. We suggested using this measure as a proxy for the difficulty of completing a particular CIA, thus related to CIA accuracy. The opinion about the validity of MODS differed. Three interviewees were positive or slightly positive to the measure, one of them claimed: "more modification clearly means it was a hard CIA. It means you couldn't immediately capture everything". On the other hand, two interviewees invalidated the measure entirely by showing that numerous modifications only deal with typos and copy-paste errors. One of them said: "I use Notepad without any spelling correction. /…/ When I paste it in the issue tracking system and submit, I get to see everything on a big screen and note the mistakes". Another interviewee instead writes the CIA directly in the issue trackers input field, but explained: "one of the problems with the tool is that it doesn't even stay open in five minutes, and it doesn't even save your data [before closing]. This is such a worst tool." When developers rush submitting CIA reports before the tool closes due to inactivity, the chances of introducing typos increases. We conclude that *MODS is not a reliable measure* for evaluating whether a solution leads to more accurate CIAs.

In conclusion, TIME appears to not at all be correlated with the time needed to conduct CIA. On the other hand, MODS received some supporters, but others show it is confounded by trivial changes to such an extent that it cannot be trusted. While both measures are easy to collect from most systems logging time stamps and revision history, they both need improvements. TIME should originate from actual changes related to an issue rather than when the issue is assigned, and MODS could possibly be filtered to remove insignificant changes such as spelling corrections.

## 5     Conclusion and Future Work

We report on an industrial case study on engineers' perspectives on Change Impact Analysis (CIA). We interviewed 14 engineers in two units of analysis in a global safety-critical software engineering context. Both the frequency of CIAs and the effort required to complete a CIA vary considerably, depending on the current phase of the development project as well as the complexity of the specific change. As a yearly average, our results suggest that developers in the case company spend roughly 50-100 hours on CIA, corresponding to one CIA per week with 1-2 hours effort each. A senior engineer also shared his rule of thumb: "CIA takes roughly 10% of the time to resolve a normal issue". We reported several major CIA challenges including communicating to developers that CIAs are necessary and beneficial, and to navigate the large document space accompanying the source code, especially the requirements. We present empirical evidence confirming that CIA is an important but costly activity in safety-critical software development, worthwhile to address in future work.

We explored engineers' attitudes by mapping their connotation of CIA versus the perceived importance of their CIAs. All combinations were identified in our study, as well as a trend that CIAs are considered increasingly important with increasing seniority. Our interviewees shared eight CIA improvement suggestions, including additional tool support, improved traceability, newcomer training, and CIA reviews. Quantitatively measuring the value of CIA support appears difficult however, as CIA is rarely conducted as an isolated activity, but rather is deeply intertwined with issue management and development in general. We found that simple analysis of CIA revisions is too confounded to evaluate improvement suggestions, thus future work is needed to develop reliable quantitative measures.

Our preliminary study is subject to a number of limitations. Due to a non-disclosure agreement, both the interviews and the analysis were done by only the first author (cf. Fig. 1). In the next step of this study, we will address the single-perspective bias by adding another researcher in a validation step, e.g., to evaluate the coding schemes. Moreover, we plan to further increase the validity of our conclusions by data triangulation, i.e., by studying real CIAs stored in the issue tracker.

When making architectural decisions on how to evolve a software system, understanding the impact of competing alternatives is important, i.e., the CIAs are valuable input to the decision maker. In prior work, we have stored trace links from previous

CIAs in semantic networks to help future developers by recommending potential impact [3, 4]. Now we aim to raise the level of abstraction to study impact of architectural decisions, i.e., to provide less granular recommendations for evolving software systems. As part of our ongoing work, we have proposed storing experiences from previous decisions in a knowledge repository [6]. As our current case study indicates that engineers put numerous hours into CIAs, and typically value their content, we argue that CIAs should also be incorporated in the knowledge repository.

**Acknowledgement.** The work is partially supported by a research grant for the ORION project (ref. number 20140218) from The Knowledge Foundation in Sweden.